# A New Asteroseismic *Kepler* Benchmark Constrains the Onset of Weakened Magnetic Braking in Mature Sun-Like Stars

Vanshree Bhalotia,[1,2] Daniel Huber,[3,4] Jennifer L. van Saders,[3] Travis S. Metcalfe,[5] Keivan G. Stassun,[6] Timothy R. White,[4] Víctor Aguirre Børsen-Koch,[7] Warrick H. Ball,[8] Sarbani Basu,[9] Aldo M. Serenelli,[10,11] Erica Sawczynec,[12] Joyce A. Guzik,[13] Andrew W. Howard,[14] and Howard Isaacson[15]

[1]*Dept. of Physics & Astronomy, University of Hawai'i, 2505 Correa Road, Honolulu, HI 96826, USA*
[2]*National Radio Astronomy Observatory,1003 Lopezville Road, Socorro, NM 87801, USA*
[3]*Institute of Astronomy, University of Hawai'i, 2680 Woodlawn Drive, Honolulu, HI 96822, USA*
[4]*Sydney Institute for Astronomy (SIfA), School of Physics, University of Sydney, NSW 2006, Australia*
[5]*White Dwarf Research Corporation, 9020 Brumm Trail, Golden, CO 80403, USA*
[6]*Department of Physics & Astronomy, Vanderbilt University, 6301 Stevenson Center Lane, Nashville, TN 37235, USA*
[7]*DARK, Niels Bohr Institute, University of Copenhagen, Jagtvej 128, 2200 Copenhagen, Denmark*
[8]*School of Physics and Astronomy, University of Birmingham, Edgbaston, Birmingham B15 2TT, UK*
[9]*Department of Astronomy, Yale University, P.O. Box 208101, New Haven, CT 06520-8101, USA*
[10]*Institute of Space Sciences (ICE, CSIC) Campus UAB, Carrer de Can Magrans, s/n, E-08193, Barcelona, Spain*
[11]*Institut d'Estudis Espacials de Catalunya (IEEC), C/Gran Capita, 2-4, E-08034, Barcelona, Spain*
[12]*Department of Astronomy, University of Texas at Austin, 2515 Speedway, Austin TX 78712-1205, USA*
[13]*Los Alamos National Laboratory, P.O. Box 1663, Los Alamos, NM 87545, USA*
[14]*Division of Physics, Mathematics and Astronomy, Caltech, 1200 E California Blvd, Pasadena CA 91125, USA*
[15]*Department of Astronomy, University of California at Berkeley, 501 Campbell Hall, Berkeley, CA 94720-3411, USA*

## Abstract

Stellar spin down is a critical yet poorly understood component of stellar evolution. In particular, results from the Kepler Mission imply that mature age, solar-type stars have inefficient magnetic braking, resulting in a stalled spin down rate. However, a large number of precise asteroseismic ages are needed for mature ($\geq$ 3Gyr) stars in order to probe the regime where traditional and stalled spin-down models differ. In this paper, we present a new asteroseismic benchmark star for gyrochronology discovered using reprocessed Kepler short cadence data. KIC 11029516 (Papayu) is a bright ($Kp =$ 9.6 mag) solar-type star with well-measured rotation period (21.1$\pm$0.8 days) from spot modulation using 4 years of Kepler long cadence data. We combine asteroseismology and spectroscopy to obtain $T_{\rm eff}$ = 5888 $\pm$ 100 K, [Fe/H] = 0.30 $\pm$ 0.06 dex, $M$ = 1.24 $\pm$ 0.05$M_\odot$, $R$ = 1.34 $\pm$ 0.02$R_\odot$ and age of 4.0 $\pm$ 0.4 Gyr, making Papayu one of the most similar stars to the Sun in terms of temperature and radius with an asteroseismic age and a rotation period measured from spot modulation. We find that Papayu sits at the transition of where traditional and weakened spin-down models diverge. A comparison with stars of similar zero-age main-sequence temperatures supports previous findings that weakened spin-down models are required to explain the ages and rotation periods of old solar-type stars.

*Keywords:* asteroseismology, gyrochronology, Kepler, magnetic braking

Corresponding author: Vanshree Bhalotia
vanshreebhalotia@gmail.com

## 1. INTRODUCTION

The search for stars with solar-like oscillations has been revolutionized in the past decades with capabilities of space telescopes such as the NASA Kepler mission (Gilliland et al. 2010a). Solar-like oscillations are convection-driven pressure waves that propagate through the interiors of stars, probing their internal structures and compositions (Chaplin & Miglio 2013; García & Ballot 2019). Fundamental stellar properties such as mass, mean density, surface gravity, and radius can be derived through global asteroseismic parameters (Huber et al. 2013; Chaplin et al. 2014; Serenelli et al. 2017). Additionally, modeling of individual oscillation frequencies yields precise stellar age measurements (Silva Aguirre et al. 2015, 2017; Metcalfe et al. 2014; Creevey et al. 2017).

Independent from asteroseismology, stellar rotation periods can also be used to predict ages of stars through gyrochronology (Barnes 2007). Gyrochronology is built upon the characteristic property of cool main sequence dwarfs to lose angular momentum as they age (Skumanich 1972). This spin down is triggered by the magnetic braking generated through the interaction of the highly-ionized stellar wind with the stellar magnetic field (Weber & Davis 1967). However, using gyrochronology as an age-estimator requires a well-calibrated relationship between the rate of stellar rotation and age, which depends predominantly on the depth of convection zone. For example, hot stars ($T_{eff}$ >6250K) beyond the Kraft break (Kraft 1967) experience minimal spin down as they have thin surface convective envelopes that are unable to sustain magnetic braking (Durney & Latour 1978). On the other hand, stars cooler than the Kraft break ($T_{eff}$ <6250K) have deeper convection zones, resulting in greater spin down. Physical changes from stellar evolution, such as increasing radii, further complicate the angular momentum loss or spin down process.

Stellar spin down has a direct impact on stellar activity. Stellar dynamo theory predicts that the magnetic field is modulated through meridional circulation, convection and differential rotation (Charbonneau 2010). Thus, since the rate of stellar rotation has significant impact on stellar activity and spot modulation (Wright et al. 2011, 2018; Santos et al. 2021; McQuillan et al. 2014), stellar activity proxies are also related to the Rossby number, which is defined as the ratio between the rotation period and the convective overturn timescale (Lehtinen et al. 2020; Noyes et al. 1984). The Rossby number is representative of the ratio between the inertial and coriolis forces within a rotating star, and appears to be a crucial metric for the effectiveness of magnetic braking.

Recent results have highlighted interesting tensions between ages from asteroseismology and those predicted by gyrochronology. Specifically, van Saders et al. (2016) found that mature-aged solar-type stars rotate faster than expected from standard spin-down models, which assumes stars spin-down at a rate inversely proportional to the square root of their age (Skumanich 1972). Additionally, limits imposed on the Rossby number, such as allowing magnetic braking to weaken after a critical Rossby number of 2.16 (hereafter $Ro_{crit}$, van Saders et al. 2016), were suggested to yield empirical agreement with expected ages. Metcalfe et al. (2016) further examined the correlation between magnetism, stellar activity and weakened magnetic braking to suggest that the stalling of such braking is caused by a decrease in the dipole component of the global magnetic field as it is disrupted by differential stellar rotation (Réville et al. 2015). More recent evidence in support of stalled spin-down include the detection a of a pile up of slowly rotating stars (David et al. 2022) and the confirmation of rotation periods using asteroseismology (Hall et al. 2021)

Despite a mounting body of evidence in its favour, a challenge for interpreting stalled spin-down models is the lack of older stars and non-solar metallicities in our gyrochronology and age calibration sample. Non-solar metallicities are relevant to the stalled spin-down hypothesis as metallicity impacts stellar convection, which is critical for spin-down (Tayar et al. 2017). Thus, our stellar rotation samples are limited and biased towards young stars in open clusters (Mamajek & Hillenbrand 2008; Meibom et al. 2015; Douglas et al. 2017; Curtis et al. 2020), and the rate at which older stars spin down is not well understood (Metcalfe & van Saders 2017). In particular, the region where magnetic braking shuts down is not well-calibrated due to a lack of solar-type stars.

While asteroseismic ages have been measured for hundreds of stars, only a small fraction have robustly measured rotation periods. For example, only 10 out of the 66 solar-like oscillators from the Kepler "Legacy" sample have rotation periods measured via spot-modulation (Silva Aguirre et al. 2017; McQuillan et al. 2014; Santos et al. 2021). The reprocessing of Kepler short-cadence data (Thompson et al. 2016) now allows a comprehensive search for additional solar-like oscillators in the Kepler data. Here we present the discovery of a new asteroseismic gyrochronology benchmark star which is similar to the Sun in terms of rotation period and age, and thus provides an excellent opportunity to test rotational spin-down models.

## 2. OBSERVATIONS

### 2.1. *Kepler Photometry*

#### 2.1.1. *Solar-like Oscillations*

For the first four quarters of the Kepler Mission, the Kepler Asteroseismic Science Consortium (KASC) surveyed ≈2000 targets, yielding ≈500 main-sequence & subgiant detections and ≈1500 non-detections of asteroseismic solar-like oscillators (Chaplin et al. 2011a). The non-detections can be partially explained by increased stellar activity (Chaplin et al. 2011a,b) or the non-optimal apertures in past Kepler data releases. Reprocessed Kepler data (DR25, Thompson et al. 2016) has already allowed new detections (Mathur et al. 2022).

To search for new solar-like oscillators, we used reprocessed data for all Kepler stars observed in short-cadence. We used stellar properties from Gaia (Berger et al. 2018) to recalculate detection probabilities using the same formalism as described in Chaplin et al. (2011a,b). Out of the 4785 stars with short-cadence data, we selected only the targets with high ($> 0.9$) Gaia-derived asteroseismic detection probabilities, pre-measured rotation periods, and a single quarter (30 days) of short-cadence data (Gilliland et al. 2010b; Berger et al. 2018; McQuillan et al. 2014). This narrowed our search down to 256 stars. We focus on one benchmark star in our sample, KIC 11029516 (hereafter Papayu[1]). Papayu was observed in long cadence for 17 quarters, and short cadence for 1 quarter via the NASA Kepler guest investigator program # GO20022 (PI Guzik). Additional new detections, including some with measured rotation periods, will be presented in Sayeed et al. (in prep). We focus on KIC11029516 as one of the highest SNR examples that allows individual frequency modeling.

### 2.1.2. *Rotation Period*

Figure 1 shows a long cadence lightcurve for one of quarters of Papayu's long-cadence data. Santos et al. (2021) reports a rotation period of $P_{rot}$ of 22.1±1.9 days, McQuillan et al. (2014) estimates 21.1±0.8 days, and a clear and consistent rotational modulation signal can be seen via visual inspection (Fig. 1) with a rotation period of ≈ 20 days. Slight differences in $P_{rot}$ estimates from Santos et al. (2021) and McQuillan et al. (2014) arise from the use of all 17 quarters and the removal of photometric pollution from nearby stars in Santos et al. (2021) compared with the use of AutoACF (McQuillan et al. 2013) in McQuillan et al. (2014). We adopt a rotation period of 21.1±0.8 days for the remainder of this study.

### 2.2. *High-Resolution Spectroscopy*

---

[1] Papayu, or પપયુ, translates to Papaya in the primary author's mother's mother tongue, Gujarati.

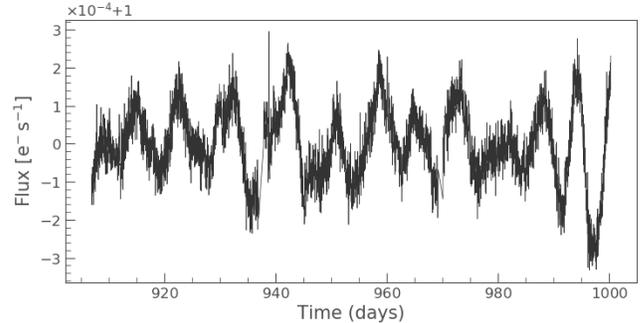

**Figure 1.** Long cadence light curve of the Quarter 10 of Kepler data for Papayu. The lightcurve is generated using the `lightkurve` package (Barentsen et al. 2020)

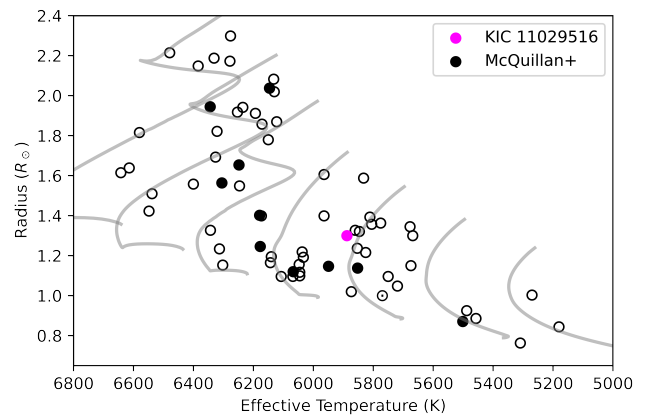

**Figure 2.** Stellar radius versus effective temperature of the Kepler Legacy sample (Silva Aguirre et al. 2017). Filled circles show stars with rotation periods determined by McQuillan et al. (2014). KIC 11029516 (Papayu) is shown in Pink.

We obtained a high-resolution spectrum of Papayu on August 14 2019 using the HIRES spectrograph (Vogt et al. 1994) at the Keck-I 10-m telescope on Maunakea observatory, Hawai'i. The spectrum was obtained and reduced as part of the California Planet Search queue (CPS, Howard et al. 2010). We obtained a 1.5 minute integration using the C5 decker, resulting in a S/N per pixel of 100 at ∼ 600 nm with a spectral resolving power of $R \sim 60000$.

To measure atmospheric parameters, we applied Specmatch-synth (Petigura 2015), which fits a synthetic grid of model atmospheres and has been extensively validated through the California Kepler Survey (Petigura et al. 2017; Johnson et al. 2017). The Specmatch-synth resulted in $T_{\rm eff} = 5888 \pm 100$ K, $\log g = 4.14 \pm 0.1$ dex, $[{\rm Fe/H}] = 0.31 \pm 0.06$ dex and $v \sin i = 3.02 \pm 1.0$ km/s. Additionally, the Specmatch-empirical parameters were $T_{\rm eff} = 5759.00 \pm 110.00$ K and [Fe/H] $= 0.31 \pm 0.09$ dex (Yee et al. 2017). Thus, these Specmatch-synth parameters are consistent with the latest Kepler-Gaia stellar properties catalog (Berger et al. 2020) and within $\approx 1$ sigma to Specmatch-empirical.

Figure 2 shows an asteroseismic HR diagram of stellar radius versus effective temperature utilizing MIST isochrones (Choi et al. 2016a). There are only ten stars out of the Kepler LEGACY sample of 66 stars with well-determined asteroseismic properties and rotation periods from spot-modulation that can be used to calibrate the relationship between gyrochronology and stellar age. Figure 2 shows Papayu is one of the most similar stars to the Sun in terms of radius and temperature with a measured rotation period from spot modulation.

### 2.3. *Bolometric Flux and Parallax*

We calculated a bolometric flux of $f_{\rm bol} = 3.68 \pm 0.11 \times 10^{-9}$ erg s$^{-1}$ cm$^{-2}$ by using 2MASS K-band photometry (Skrutskie et al. 2006) and applying a bolometric correction from MIST isochrones (Choi et al. 2016b), as implemented in `isoclassify` (Huber et al. 2017). We used a bolometric correction error floor of 0.03 mag, consistent with the expected systematic offsets for bolometric fluxes (Zinn et al. 2019; Tayar et al. 2020). Interstellar extinction from 3D dustmaps (Green et al. 2015) was found to be negligible, consistent with the short distance ($\approx 130$ pc). Finally, we combined our $f_{\rm bol}$ value with the Gaia DR3 parallax (Lindegren et al. 2021) to calculate luminosity of $1.91 \pm 0.05 \, L_\odot$.

As an independent check we performed a spectral energy distribution (SED) fit following the procedures described in Stassun & Torres (2016); Stassun et al. (2017, 2018). We used $B_T V_T$ magnitudes from *Tycho-2*, $JHK_S$ magnitudes from *2MASS*, W1–W4 magnitudes from *WISE*, $G_{\rm BP} G_{\rm RP}$ magnitudes from *Gaia*, the NUV magnitude from *GALEX*, and the *U* magnitude from the KIS survey. We performed a fit using PHOENIX stellar atmosphere models (Husser et al. 2013), with the free parameters being the effective temperature ($T_{\rm eff}$) and the extinction $A_V$, which we limited to maximum line-of-sight value from the Galactic dust maps of Schlegel et al. (1998). We also adopted the metallicity ([Fe/H]) from the spectroscopic analysis above. Integrating the (unreddened) best-fit model SED gives the bolometric flux at Earth, $F_{\rm bol} = 3.873 \pm 0.090 \times 10^{-9}$ erg s$^{-1}$ cm$^{-2}$. Combining $F_{\rm bol}$ with the *Gaia* parallax gives the luminosity $L_{\rm bol} = 1.998 \pm 0.047$ L$_\odot$, in good agreement with the value derived using bolometric corrections.

## 3. ASTEROSEISMIC DATA ANALYSIS

### 3.1. *Data Preparation*

We used the single quarter of available short-cadence data for our analysis. PDCSAP flux was preferred over SAP flux, as it has been detrended for any systematic effects (Smith et al. 2012; Stumpe et al. 2012). We ignored data points with non-zero quality flags and applied a 3-sigma clipping of the lightcurve with a box width of 50 datapoints to remove any outliers. Furthermore, to focus on the high frequency asteroseismic oscillations, we removed data artifacts due to *Kepler* observing quarters and normalized the lightcurve using a box filter of 0.05 days to remove long periodic variations caused by instrumental effects (thermal variations, drifts) and/or stellar varibility (stellar rotation, stellar activity). The power spectrum generated after making these corrections is shown in Figure 3, which shows a clear detection of solar-like oscillations around 2000 $\mu$Hz.

### 3.2. *Background Modeling*

In order to separate the asteroseismic oscillations from the granulation signal and background noise, we used the Bayesian modelling software package DIAMONDS (Corsaro & De Ridder 2014). We used a composite stellar background model which consisted of constant photon noise, stellar activity variations, a sampling response function and two Harvey models (Harvey 1985). Eight total parameters were involved in the fit, including the white noise threshold (in ppm$^2$/$\mu$Hz), the Harvey model parameters with a characteristic frequency ($\mu$Hz) and amplitude (ppm), along with the frequency of maximum power ($\nu_{max}$), the height (in ppm$^2$/$\mu$Hz) and the FWHM of the power excess.

Figure 3 shows the fitted background model and its individual components calculated via DIAMONDS. Initial guesses were estimated through visual inspection of the power spectrum between 270 and 8333 $\mu$Hz. The goodness and convergence of fit was assessed through DIAMONDS via the posterior probability distributions (Appendix A). The final values for the background fit, calculated as the median and 1-$\sigma$ confidence limits of the posterior distributions, are listed in Table 1.

### 3.3. *Individual Frequencies*

After correcting the power spectrum for the background, we extracted individual frequencies using the Peakbagging module of DIAMONDS (Corsaro & De Ridder 2014). The module operates by generating a set of multiple Lorentzian profiles. Similar to the Backgrounds module, the PeakBagging module requires initial guesses of each peak's linewidth or $\Gamma_i$ ($\mu$Hz), amplitude or $A_i$ (ppm) and central frequency or $\nu_{0,i}$ ($\mu$Hz). We identified 15 possible peaks based on manual cross-identification. Thus, 45 initial guesses ($\Gamma_i$, $A_i$ & $\nu_{0,i}$ for

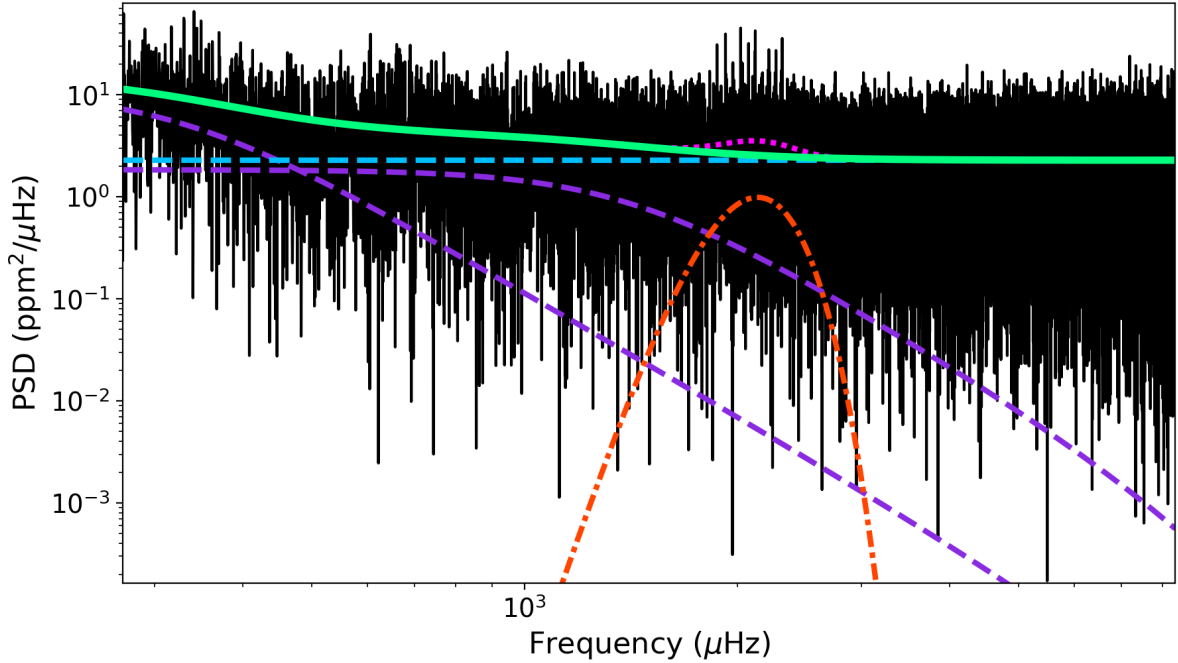

**Figure 3.** Power spectrum of the single quarter of short-cadence data for Papayu after outlier correction and high-pass filtering. The solid green line is the modeled granulation background. The blue dashed line is a constant white noise, the purple dashed curves are the Harvey models, the dash-dotted orange curve is the height of the oscillation and the dotted pink curve is the frequency of maximum power and the FWHM of the oscillation.

**Table 1.** Model Parameters of the Background fit

| Parameter | Median value |
|---|---|
| w ($ppm^2/\mu$Hz) | $2.28^{+0.02}_{-0.01}$ |
| $\sigma_{long}$ (ppm) | $62.17^{+3.27}_{-2.70}$ |
| $b_{long}$ ($\mu$Hz) | $322.14^{+19.01}_{-25.33}$ |
| $\sigma_1$ (ppm) | $53.12^{+1.47}_{-1.17}$ |
| $b_1$ ($\mu$Hz) | $1,378.86^{+170.40}_{-201.89}$ |
| $H_{osc}$ ($ppm^2/\mu$Hz) | $1.04^{+0.10}_{-0.11}$ |
| $\nu_{max}$ ($\mu$Hz) | $2,141.18^{+29.81}_{-28.74}$ |
| $\sigma_{env}$ ($\mu$Hz) | $241.58^{+44.79}_{-49.56}$ |

Notes: white noise (w), $1^{st}$ Harvey model amplitude ($\sigma_{long}$), $1^{st}$ harvey model turnoff ($b_{long}$), $2^{nd}$ harvey model amplitude ($\sigma_1$), $2^{nd}$ harvey model turnoff ($b_1$), height of oscillation envelope ($H_{osc}$), frequency of maximum oscillation ($\nu_{max}$) & FWHM of oscillation envelope ($\sigma_{env}$)

$0 < i \leq 15$) with symmetric upper and lower limits were provided to DIAMONDS in a PeakBagging run.

We performed mode identification of extracted frequencies using an échelle diagram (Grec et al. 1983). An échelle diagram is a compact visualization of a power spectrum, as it breaks the power spectrum up into $\Delta\nu$-sized sections (radial orders) and stacks them atop each other. We expect an échelle diagram to show vertical ridges of quadrupole ($l=2$), radial ($l=0$) and dipole ($l=1$) modes.

The initial guesses for linewidths and amplitudes were based on visual inspection and then tuned by fitting only a few sets of peaks at a time. Cumulative addition of the peaks allowed tuning of the linewidth and amplitude guesses at each step. Figure 4 shows the fit to the background-corrected power spectrum, with Lorentzian profiles derived from these 45 parameters. The final values and posteriors of all 45 parameters are shown in Table 2 and Fig. 11.

Figure 6 shows an échelle diagram of the extracted frequencies. We successfully extracted 5 radial modes, 6 dipole modes and 4 quadruple modes. The value of large frequency separation, $\Delta\nu$, was measured to be $97.31 \pm 0.14$ $\mu$Hz by taking the mean separation between consecutive radial modes. Similarly, using the mean of the small frequency separations for each radial order, we calculated a $\delta\nu_{02}$ value of $6.12 \pm 0.15$ $\mu$Hz.

## 4. ASTEROSEISMIC MODELING

### 4.1. *Scaling relations and the C-D Diagram*

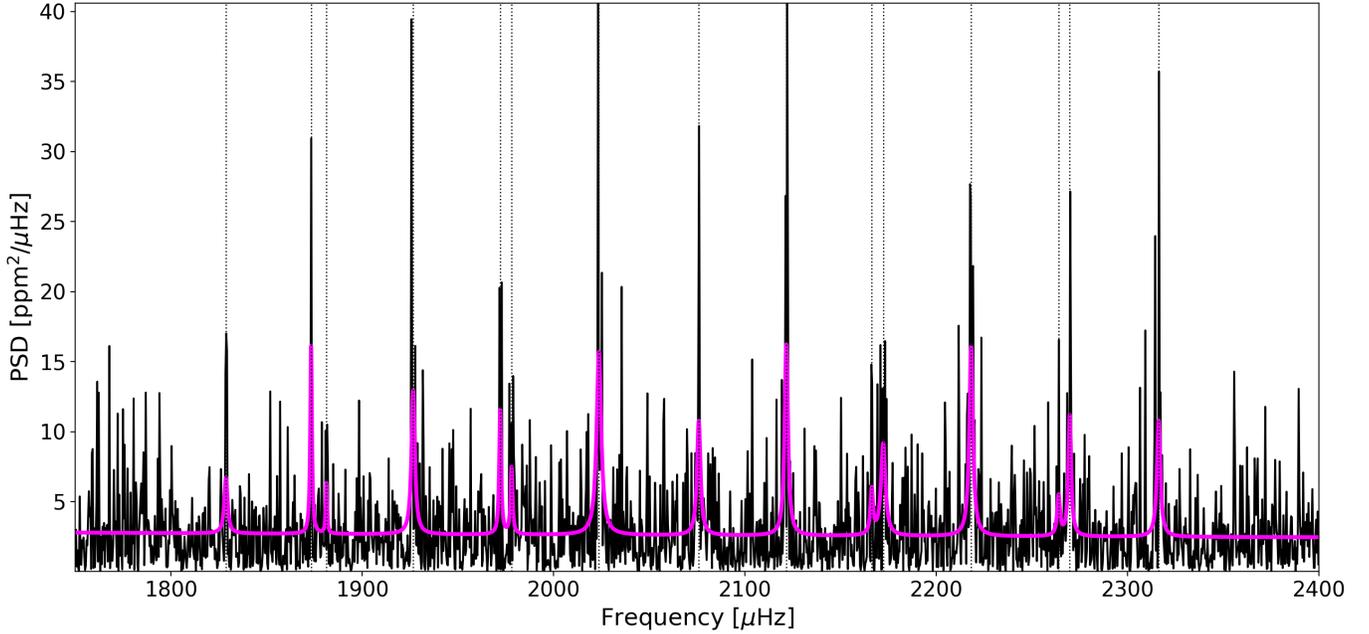

**Figure 4.** Identified oscillation frequencies for Papayu. The Background-subtracted Power Spectrum is in black, the Lorentzian profiles that fit each of the 15 oscillation frequencies are in pink. Posteriors and final values for the parameters used in this fit are given in Appendix A.

**Table 2.** Best-fit values for Papayu's oscillation frequencies.

| Peak # | $n_i$ | $l_i$ | $\nu_{0,i}$ ($\mu$Hz) | $A_i$ (ppm) | $\Gamma_i$ ($\mu$Hz) |
|---|---|---|---|---|---|
| 1  | 17 | 1 | $1{,}828.73^{+0.23}_{-0.25}$ | $5.13^{+0.66}_{-0.64}$   | $2.02^{+0.53}_{-0.59}$ |
| 2  | 17 | 2 | $1{,}873.23^{+0.10}_{-0.11}$ | $6.73^{+0.72}_{-0.78}$   | $1.03^{+0.40}_{-0.33}$ |
| 3  | 18 | 0 | $1{,}881.22^{+0.29}_{-0.33}$ | $3.07^{+0.41}_{-0.42}$   | $0.80^{+0.17}_{-0.17}$ |
| 4  | 18 | 1 | $1{,}926.60^{+0.28}_{-0.28}$ | $8.46^{+0.77}_{-0.73}$   | $2.12^{+0.37}_{-0.42}$ |
| 5  | 18 | 2 | $1{,}972.06^{+0.21}_{-0.14}$ | $5.93^{+0.69}_{-0.60}$   | $1.21^{+0.28}_{-0.30}$ |
| 6  | 19 | 0 | $1{,}978.00^{+0.26}_{-0.27}$ | $4.42^{+0.40}_{-0.38}$   | $1.25^{+0.24}_{-0.24}$ |
| 7  | 19 | 1 | $2{,}023.67^{+0.20}_{-0.24}$ | $11.16^{+0.73}_{-0.91}$  | $2.88^{+0.38}_{-0.40}$ |
| 8  | 20 | 0 | $2{,}075.85^{+0.15}_{-0.18}$ | $7.73^{+0.56}_{-0.60}$   | $2.21^{+0.51}_{-0.47}$ |
| 9  | 20 | 1 | $2{,}121.61^{+0.14}_{-0.15}$ | $9.75^{+0.62}_{-0.54}$   | $2.10^{+0.42}_{-0.48}$ |
| 10 | 20 | 2 | $2{,}166.26^{+0.19}_{-0.17}$ | $4.13^{+0.43}_{-0.45}$   | $1.62^{+0.25}_{-0.24}$ |
| 11 | 21 | 0 | $2{,}172.25^{+0.38}_{-0.31}$ | $7.73^{+0.57}_{-0.52}$   | $2.73^{+0.44}_{-0.48}$ |
| 12 | 21 | 1 | $2{,}218.20^{+0.16}_{-0.16}$ | $10.17^{+0.58}_{-0.54}$  | $2.29^{+0.29}_{-0.36}$ |
| 13 | 21 | 2 | $2{,}263.91^{+0.26}_{-0.27}$ | $4.04^{+0.58}_{-0.66}$   | $1.73^{+0.25}_{-0.31}$ |
| 14 | 22 | 0 | $2{,}269.76^{+0.16}_{-0.17}$ | $7.13^{+0.53}_{-0.45}$   | $1.76^{+0.22}_{-0.25}$ |
| 15 | 22 | 1 | $2{,}316.22^{+0.20}_{-0.19}$ | $7.71^{+0.49}_{-0.51}$   | $2.14^{+0.25}_{-0.17}$ |

Column names denote the Peak Number, $i$ (increasing from left to right in Fig. 4), Radial Order ($n_i$), Spherical Degree ($l_i$), Central peak frequency ($\nu_{0,i}$), Peak amplitude ($A_i$), & Peak line width ($\Gamma_i$)

We calculated a preliminary asteroseismic age using an asteroseismic HR diagram (hereafter C-D diagram Christensen-Dalsgaard 1984) and asteroseismic scaling relationships (Bellinger 2019). Physically, the large frequency separation scales with the square root of the mean stellar density, while the small separation scales with the sound speed gradient in the core (Ulrich 1986). Thus the small separation is sensitive to stellar age on the main sequence.

Figure 5 shows a C-D diagram of $\delta\nu_{02}$ against $\Delta\nu$ for Papayu's metallicity ([Fe/H] = 0.30), with stellar-evolution models from ASTEC & ADIPLS (Christensen-Dalsgaard 2007a, 2008) as described in White et al. (2011). In the C-D diagram, stars evolve from the top right to the bottom left. The range of stellar masses in the C-D diagram is from 1.0 to 1.6 $M_\odot$, with the latter end slightly above the Kraft break (Kraft 1967). Using Papayu's values of $\Delta\nu$ and $\delta\nu_{02}$ from Sec. 3 we are able to constrain Papayu's age from the C-D diagram to $\approx 4.0$ Gyr. Papayu's age can also be measured by substituting its stellar properties from Table 3 into the age-scaling relationship in Bellinger (2019). From this we obtain an age estimate of $3.64 \pm 0.31$ Gyr.

### 4.2. *Frequency Modeling*

To derive a more robust age, we modeled Papayu using the stellar evolution code MESA (Paxton et al. 2011, 2013, 2015, 2018, 2019) using effective temperature, metallicity, log g, $\Delta\nu$ ($\mu$Hz), $\delta\nu_{02}$ (Table 3) and all the frequencies obtained using the DIAMONDS module (Table 2). All observed frequencies and parameters were added to a MESA inlist. This inlist uses the `astero` and `GYRE` modules (Townsend & Teitler 2013) to generate a pre-main sequence model and perform a `simplex` search. The `simplex` search executes multiple runs in order to obtain the stellar model with the best fit to the observed parameters and frequencies. While the `astero` module obtains the best fits for each parameter, the linked `GYRE` stellar oscillation module computes modelled frequencies using a parallelized multiple Magnusson shooting scheme.

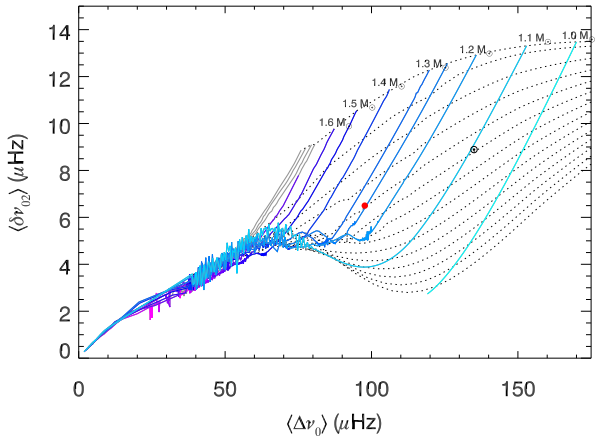

**Figure 5.** A C-D diagram, with Papayu's position shown in red and the Sun's position shown in black. The solid lines are stellar evolution models tailored to Papayu's metallicity of 0.30 dex, therefore the Sun does not cross the 1 $M_\odot$ model. Stars on these model tracks evolve from the top right to the bottom left. The dashed lines are isochrones from zero to 12 Gyr (top to bottom) in steps of 2 Gyr.

Specifically, the input physics was selected for variable mixing length parameter ($\alpha$), helium abundance (Y), metallicity ([Fe/H]) and mass. Minimum and maximum thresholds for these varying parameters and initial values were provided. The target value and error in log g was computed through asteroseismic scaling relations, similar to the initial guess for mass. The initial guesses for effective temperature and metallicity were provided through spectroscopic observations as described in Section 2.2. Initial values for the helium abundance and mixing length were 0.27±0.03 and 1.8±0.2 respectively. A cubic surface correction scheme, utilizing the default correction factor from Ball, W. H. & Gizon, L. (2017), is applied to adjust for the systematic disparity between modeled and observed mode frequencies due to inaccuracies in modeling near-surface stellar layers. The temperature structure of the atmosphere was defined using the Eddington grey relation, with opacity varying with optical depth. Elemental diffusion is allowed, and a radiative turbulence coefficient of unity is selected in accordance with Morel, P. & Thévenin, F. (2002). For varying helium abundance and mixing length, we performed runs varying overshooting, keeping it fixed to the default value of 0.015 and keeping it fixed to 0, but found no significant differences in MESA outputs. Papayu is on the verge of developing a convective core at 1.24 ± 0.05 $M_\odot$, so it appears overshooting is not a significant concern yet (Claret & Torres 2018). Additionally, no value for $\nu_{max}$ was provided during our runs, and the $\chi^2$ seismic fraction was increased to 0.8, compared to the default value of 0.667. Partial evaluation controls, chi-squared based time-step controls and stopping conditions were slightly altered to fine-tune our measurements, and all other parameters were unchanged.

MESA outputs multiple runs, each with frequencies and stellar parameters. Frequency and stellar parameter outputs were plotted with respect to run number to ensure compliance with input physics. Observed and modelled frequencies for the best-fitting model are shown in Figure 7, demonstrating good agreement. Stellar parameter values along with their corresponding statistical uncertainties for the best-fit MESA model are as follows: Mass = 1.23 ± 0.01 $M_\odot$, Radius = 1.34 ± 0.01 $R_\odot$, log $g$ = 4.276 ± 0.001 cgs, Density = 0.728 ± 0.004 gcc, and Age = 4.0 ± 0.2 Gyr. To test for systematic errors, multiple coauthors modeled observed oscillation frequencies using other stellar evolution codes and modeling methods, including BeSPP, YREC, BASTA, MESA and AMP (Demarque et al. 2007; Christensen-Dalsgaard 2007b; Metcalfe et al. 2009; Paxton et al. 2011, 2013, 2015, 2018, 2019; Ball, W. H. & Gizon, L. 2017; Serenelli et al. 2017; Aguirre Børsen-Koch et al. 2022). Model inputs include individual frequencies (Table 2), spectroscopic (Keck/HIRES) $T_{eff}$ and [Fe/H], along with $\Delta\nu$, $\nu_{max}$, and Gaia-derived luminosity. Modeling efforts yielded consistent results, both with MESA derived parameters and outputs from various pipelines. We adopted the best-fit values from MESA modeling described above, with uncertainties estimated by adding in quadrature the formal uncertainty from MESA modeling and the standard deviation of the values over all methods. The final adopted values are listed in Table 3. Our final age uncertainty is 10%, and the value agrees within 4% to alternative estimates described in Section 4.1. Furthermore, Papayu's Rotation Period and Age align with the green M67 isochrone depicted in Figure 4 of Barnes et al. (2016). This suggests the gyrochrone age of a star exhibiting Papayu's rotation period corresponds to the age we computed in this section.

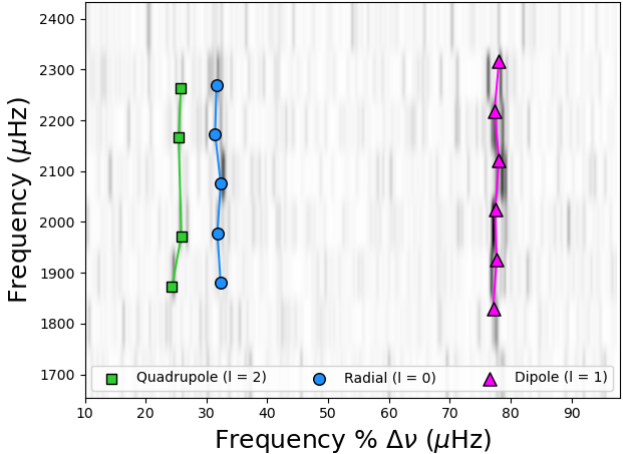

**Figure 6.** Échelle diagram using frequencies extracted by the PeakBagging module of the DIAMONDS package. The colours correspond to the different degrees of spherical harmonics as denoted in the legend, with the radial mode denoted in blue, the dipole mode denoted in pink, and the quadruple mode denoted in green. The échelle diagram is created via `echelle` (Hey & Ball 2020).

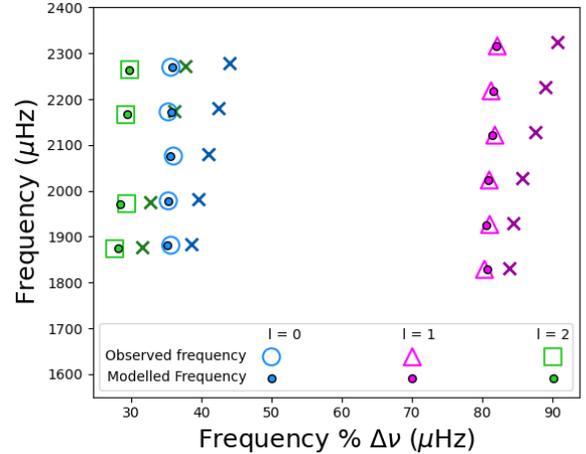

**Figure 7.** Échelle diagram comparing observed frequencies (unfilled shapes) to frequencies obtained from our best-fit MESA module (filled dots and crosses). The colours of the symbols represent the various spherical degrees, with the radial mode in blue, dipole mode in pink, and quadruple mode in green. Original MESA derived frequencies are denoted by coloured crosses, and the surface-corrected MESA frequencies are shown in filled dots.

### 4.3. *Comparison to Age from Chromospheric Activity*

The Keck/HIRES spectrum (Section 2.2) allows a measurement of the chromospheric activity from Ca H&K lines. We measured $\log R'_{HK} = -5.061$ using the method of Isaacson & Fischer (2010). After correcting for the metallicity, we calculated $\log R'_{HK}[T_{eff}] = -4.968$ on the scale of Lorenzo-Oliveira et al. (2018). The corresponding age inferred from chromospheric activity is $3.9^{+1.1}_{-0.9}$ Gyr, with an error bar estimated from the uncertainty in the age-activity calibration. The estimated age from chromospheric activity provides a good match to the asteroseismic age of $4.0 \pm 0.4$ Gyr.

## 5. DISCUSSION

Papayu's robust rotation period and asteroseismic age allows us to place the star in context with other known asteroseismic gyrochronology benchmarks. Figure 8 shows rotation period as a function of stellar age for all asteroseismic gyrochronology benchmark stars in the LEGACY sample along with Metcalfe et al. (2016) and Hall et al. (2021). The stars in Figure 8 provide a comprehensive overview of known asteroseismic gyrochronology rotators. Almost all the stars in this sample have asteroseismic ages and fundamental stellar properties through Silva Aguirre et al. (2017) & Silva Aguirre et al. (2015). There are a handful of stars that did not have ages or stellar properties measured in both LEGACY surveys but were found in Metcalfe et al. (2016). The rotation periods of stars are obtained either through rotational modulation measurements (McQuillan et al. 2014; García et al. 2014) or through asteroseismic splitting measurements (Hall et al. 2021; Davies et al. 2015). Rotation periods measured via spot modulation were preferred over asteroseismic splittings, with García et al. (2014) being given preference over McQuillan et al. (2014). When periods were obtained through asteroseismic splittings, Davies et al. (2015) was given preference over Hall et al. (2021).

**Table 3.** Stellar Parameters for Papayu

| | |
|---|---:|
| **Star Information** | |
| KIC ID | 11029516 |
| Tycho ID | TYC 3547-1118-1 |
| 2MASS ID | J19270298+4835118 |
| Coordinates (ICRS) | 19h 27m 02s, +48° 35' 11" |
| Kepler magnitude (mag) | 9.6 |
| **Observables from Kepler** | |
| $\nu_{max}(\mu\mathrm{Hz})$ | $2121.7 \pm 0.1$ |
| $\Delta\nu(\mu\mathrm{Hz})$ | $97.1 \pm 0.1$ |
| $\delta\nu_{02}(\mu\mathrm{Hz})$ | $6.1 \pm 0.2$ |
| Rotation Period (d) | $21.1 \pm 0.8$ |
| **Photometry & Gaia** | |
| $f_{\mathrm{bol}}(10^{-9}\,\mathrm{erg\,s^{-1}\,cm^{-2}})$ | $3.68 \pm 0.11$ |
| Luminosity$(L_\odot)$ | $1.91 \pm 0.05$ |
| **Spectroscopy** | |
| $T_{\mathrm{eff}}$ (K) | $5888 \pm 100$ |
| [Fe/H](dex) | $0.30 \pm 0.06$ |
| **Asteroseismic modeling** | |
| Mass ($M_\odot$) | $1.23 \pm 0.03$ |
| Radius ($R_\odot$) | $1.34 \pm 0.01$ |
| $\log g$(cgs) | $4.28 \pm 0.01$ |
| Density (gcc) | $0.728 \pm 0.005$ |
| Age (Gyr) | $4.0 \pm 0.4$ |

The stars in Figure 8 are coloured by their Zero Age Main Sequence (ZAMS) temperatures. The ZAMS temperature is a function of mass and metallicity and selects stars with similar structures and convection zone depths. Convective overturn timescales, which scale with convection zone depth, enter in the Rossby number and are therefore an important parameter in braking laws. The ZAMS temperatures were obtained by interpolating model grids for a given stellar mass and metallicity using the `kiauhoku` package (Claytor et al. 2020).

Figure 8 shows that Papayu is similar to the Sun in terms of rotation and age for stars with known spot modulation measurements. Figure 9 takes a closer look into this smaller subset of Figure 8, displaying the period-age space as a function of ZAMS temperatures for a sub-sample of asteroseismic gyrochronology benchmark stars with well-constrained rotation periods (with errors ≤ 10%) and ZAMS temperatures within 100 K of Papayu's. We also eliminated stars from Hall et al. (2021) which have rotation period uncertainties larger than 10%.

In order to ascertain whether Papayu experiences weakened magnetic braking, we calculated a spin down track for the average star in this sub-sample. To construct these tracks, we first averaged the ZAMS temperature amongst all the stars in the subset. We then calculated a model grid using weakened magnetic braking for the spin down for a fixed solar mass (1 $M_\odot$) and varied the metallicity (0.2 dex) until we reached a model with ZAMS temperature that matches the observed ZAMS temperature of the sample. The $Ro_{crit}$ model grid was computed via `kiauhoku` (Claytor et al. 2020), which uses the modified magnetic braking law based on van Saders & Pinsonneault (2013), taking into account stalled stellar spin-down based on a critical Rossby number of 2.16 (van Saders et al. 2016). We also calculated standard spin down models using the fast launch conditions of van Saders & Pinsonneault (2013) anchored to the rapidly rotating envelope in the Pleiades and M37 open clusters, and do not take into account the effect of a critical Rossby number for magnetic braking. It is worth noting that assumptions about the initial rotation period do not significantly shift the rotational model tracks (van

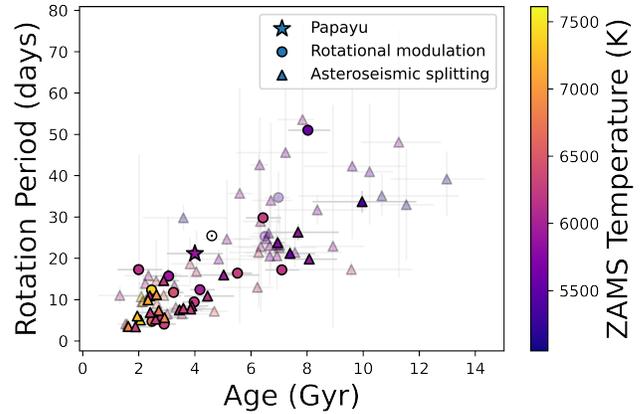

**Figure 8.** Rotation period versus asteroseismic age for known asteroseismic gyrochronology benchmark stars color-coded by ZAMS temperature. Rotation periods are obtained through stellar modulation (circles) or asteroseismic frequency splittings (triangles). Fainter points show rotation period measurements with uncertainties exceeding 10%.

Saders & Pinsonneault 2013; van Saders et al. 2016; Metcalfe et al. 2020), and that metal poor stars spin down more effectively than metal rich stars in the early phases of spin down (Amard & Matt 2020). However, in later spin down phases this reverses, and metal poor stars spin down less effectively than metal rich stars. The `kiauhoku` model grids take this effect into account (Claytor et al. 2020).

Figure 9 shows the Papayu sits at the transition region between both Skumanich-like and $Ro_{crit}$ model tracks, thus, it provides an anchor point where both spin down model tracks diverge. We note that the models do provide a good match for the Sun since they trace the average mass and metallicity for a star in this sample. We observe that the $Ro_{crit}$ or stalled spin down model provides a significantly better match to the sample, confirming previous results by implying a preference for the weakened magnetic braking law.

## 6. CONCLUSIONS

We have presented the discovery of a new asteroseismic gyrochronology benchmark star in the Kepler field (KIC11029516 or Papayu). Our main conclusions are as follows:

- Papayu was discovered using Kepler short-cadence data post-DR25 reprocessing, and is a bright (V = 9.63 mag) solar-type star, with a spectroscopic $T_{eff}$ of 5888 ± 100 K and [Fe/H] of 0.30 ± 0.06 dex. We used DIAMONDS frequency-modeling in conjunction with MESA-modelling to obtain its mass of 1.23 ± 0.03 $M_\odot$, radius of 1.34 ± 0.01 $R_\odot$

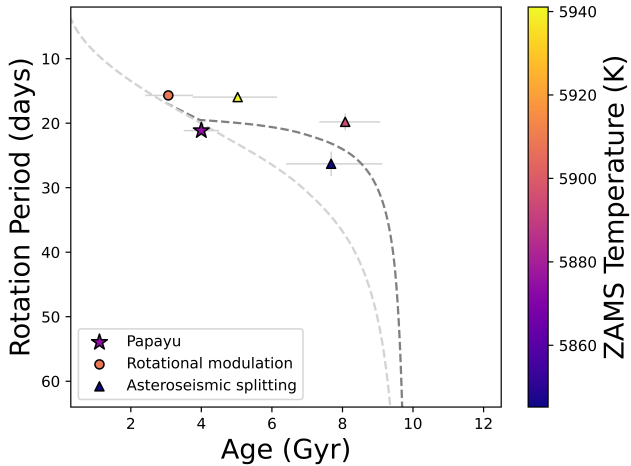

**Figure 9.** Rotation period versus asteroseismic age for a subset of asteroseismic gyrochronology benchmark stars within 100 K of Papayu's ZAMS temperature and with rotation period measurement errors less that 10%. These stars correspond to the brighter coloured ones in Fig. 8 and are similarly coloured by their ZAMS temperatures, which fall in a range $\leq$ 5974K or $\geq$ 5774K, a symmetric interval around Papayu's ZAMS temperature of 5874K. The dark grey line corresponds to `kiauhoku`'s $Ro_{crit}$ model, and the light gray line corresponds to `kiauhoku`'s fastlaunch model for a mass and metallicity that reproduced the mean ZAMS temperature of the sample.

and most importantly, asteroseismic age of 4.0 ± 0.4 Gyr.

- Among oscillating Kepler stars, Papayu is one of the most similar stars to the Sun in terms of asteroseismic age and rotation period. 17 quarters of long-cadence data make it possible to use rotational modulation to obtain its rotation period of 21.1±0.8 days (McQuillan et al. 2014)

- Papayu is an asteroseismic gyrochronology benchmark anchoring the onset of weakened magnetic braking. It is at the transition where both spin-down models diverge, and a comparison with stars of similar ZAMS temperatures supports previous findings that weakened spin-down models are required to explain the ages and rotation periods of old solar-type stars. Its higher-than average metallicity should allow tests of the composition dependence of the spin down laws.

The discovery of Papayu shows that several exciting asteroseismic gyrochronology benchmarks are still waiting to be discovered in the Kepler dataset. Recent analyses of the Kepler asteroseismic survey sample have already increased the Kepler yield (Mathur et al. 2022), and upcoming efforts will provide the first homogeneous analysis of the entire Kepler asteroseismic short-cadence dataset (Sayeed et al., in prep). Furthermore, results from K2 and TESS (Metcalfe et al. 2020; Hatt et al. 2023) will continue to contribute to our understanding of the connections between stellar age, activity, and rotation. Lastly, as discussed in Saunders et al. (2024), we emphasize there is uncertainty on $\approx$ Gyr scales regarding the exact point at which solar-like stars begin weakened braking. This is primarily driven by the lack of calibrating sources around the regime where weakened magnetic braking models diverge from the standard. The process of precisely anchoring the onset of weakened magnetic braking starts with gathering various Papayu-like evolutionary statistics, with this study marking the beginning of this endeavour.

We gratefully acknowledge everyone involved in the *Kepler* mission, the W.M. Keck Observatory & STScI for their efforts, which have made this paper possible. The authors wish to recognize and acknowledge the very significant cultural role and reverence that the summit of Maunakea has always had within the indigenous Hawaiian community. We are most fortunate to have the opportunity to conduct observations from this mountain. The primary author wishes to greet Maunakea with पञ्चाङ्ग प्रणाम, offering utmost reverence to the piko hoʻokahi and her Kānaka Maoli descendants. These astronomical observations would not have been possible without Maunakea's haʻawina. Guided by pono and our कर्त्तव्य, we use the discovery of पपयु to malama this ʻāina and her kānaka. V.B. thanks her parents, siblings, mentors, and आदि पराशक्ति for their आशीर्वाद during this project that has culminated in her first first-author scientific publication.

V.B. and D.H. acknowledge support from the National Aeronautics and Space Administration (80NSSC19K0597). D.H. also acknowledges support from the Alfred P. Sloan Foundation and the Australian Research Council (FT200100871). JvS acknowledges support from the National Aeronautics and Space Administration (80NSSC19K0597). SB acknowledges NSF grant AST-2205026. T.S.M. acknowledges NASA grant 80NSSC22K0475. Computational time at the Texas Advanced Computing Center was provided through XSEDE allocation TG-AST090107. AMS acknowledges grants PID2019-108709GB-I00 from Ministry of Science and Innovation (MICINN, Spain), Spanish program Unidad de Excelencia María de Maeztu CEX2020-001058-M, 2021-SGR-1526 (Generalitat de Catalunya), and support from ChETEC-INFRA (EU project no. 101008324). All the photometric data presented in this paper was obtained from the Mikulski Archive for Space Telescopes (MAST). STScI is operated by the Association of Universities for Research in Astronomy, Inc., under NASA contract NAS5-26555. Support for MAST for non-HST data is provided by the NASA Office of Space Science via grant NNX13AC07G and by other grants and contracts. The spectral data presented herein were obtained at the W. M. Keck Observatory, which is operated as a scientific partnership among the California Institute of Technology, the University of California and the National Aeronautics and Space Administration. The Observatory was made possible by the generous financial support of the W. M. Keck Foundation.

*Software:* lightkurve (Barentsen et al. 2020), echelle (Hey & Ball 2020), DIAMONDS (Corsaro & De Ridder 2014), kiauhoku (Claytor et al. 2020), MESA (Paxton et al. 2011, 2013, 2015, 2018, 2019), Astropy (Astropy Collaboration et al. 2018), Matplotlib (Hunter 2007), NumPy (Oliphant 2006; Van Der Walt et al. 2011) and Pandas (pandas development team 2020)

New Asteroseismic Gyrochronology Benchmark with Kepler 13Serenelli, A., Johnson, J., Huber, D., et al. 2017, ApJS, 233, 23, doi: 10.3847/1538-4365/aa97df

Silva Aguirre, V., Davies, G. R., Basu, S., et al. 2015, MNRAS, 452, 2127, doi: 10.1093/mnras/stv1388

Silva Aguirre, V., Lund, M. N., Antia, H. M., et al. 2017, ApJ, 835, 173, doi: 10.3847/1538-4357/835/2/173

Skrutskie, M. F., Cutri, R. M., Stiening, R., et al. 2006, AJ, 131, 1163, doi: 10.1086/498708

Skumanich, A. 1972, ApJ, 171, 565, doi: 10.1086/151310

Smith, J. C., Stumpe, M. C., Cleve, J. E. V., et al. 2012, Publications of the Astronomical Society of the Pacific, 124, 1000, doi: 10.1086/667697

Stassun, K. G., Collins, K. A., & Gaudi, B. S. 2017, AJ, 153, 136, doi: 10.3847/1538-3881/aa5df3

Stassun, K. G., Corsaro, E., Pepper, J. A., & Gaudi, B. S. 2018, AJ, 155, 22, doi: 10.3847/1538-3881/aa998a

Stassun, K. G., & Torres, G. 2016, AJ, 152, 180, doi: 10.3847/0004-6256/152/6/180

Stumpe, M. C., Smith, J. C., Cleve, J. E. V., et al. 2012, Publications of the Astronomical Society of the Pacific, 124, 985, doi: 10.1086/667698

Tayar, J., Claytor, Z. R., Huber, D., & van Saders, J. 2020, arXiv e-prints, arXiv:2012.07957. https://arxiv.org/abs/2012.07957

Tayar, J., Somers, G., Pinsonneault, M. H., et al. 2017, ApJ, 840, 17, doi: 10.3847/1538-4357/aa6a1e

Thompson, S. E., Caldwell, D. A., Jenkins, J. M., et al. 2016, Kepler Data Release 25 Notes (KSCI-19065-002)

Townsend, R. H. D., & Teitler, S. A. 2013, Monthly Notices of the Royal Astronomical Society, 435, 3406, doi: 10.1093/mnras/stt1533

Ulrich, R. K. 1986, ApJL, 306, L37, doi: 10.1086/184700

Van Der Walt, S., Colbert, S. C., & Varoquaux, G. 2011, Computing in Science & Engineering, 13, 22

van Saders, J. L., Ceillier, T., Metcalfe, T. S., et al. 2016, Nature, 529, 181, doi: 10.1038/nature16168

van Saders, J. L., & Pinsonneault, M. H. 2013, ApJ, 776, 67, doi: 10.1088/0004-637X/776/2/67

Vogt, S. S., Allen, S. L., Bigelow, B. C., et al. 1994, in Society of Photo-Optical Instrumentation Engineers (SPIE) Conference Series, Vol. 2198, Instrumentation in Astronomy VIII, ed. D. L. Crawford & E. R. Craine, 362, doi: 10.1117/12.176725

Weber, E. J., & Davis, Leverett, J. 1967, ApJ, 148, 217, doi: 10.1086/149138

White, T. R., Bedding, T. R., Stello, D., et al. 2011, The Astrophysical Journal, 743, 161, doi: 10.1088/0004-637x/743/2/161

Wright, N. J., Drake, J. J., Mamajek, E. E., & Henry, G. W. 2011, ApJ, 743, 48, doi: 10.1088/0004-637X/743/1/48

Wright, N. J., Newton, E. R., Williams, P. K. G., Drake, J. J., & Yadav, R. K. 2018, MNRAS, 479, 2351, doi: 10.1093/mnras/sty1670

Yee, S. W., Petigura, E. A., & von Braun, K. 2017, ApJ, 836, 77, doi: 10.3847/1538-4357/836/1/77

Zinn, J. C., Pinsonneault, M. H., Huber, D., et al. 2019, ApJ, 885, 166, doi: 10.3847/1538-4357/ab44a9

14      Bhalotia et al.

## APPENDIX

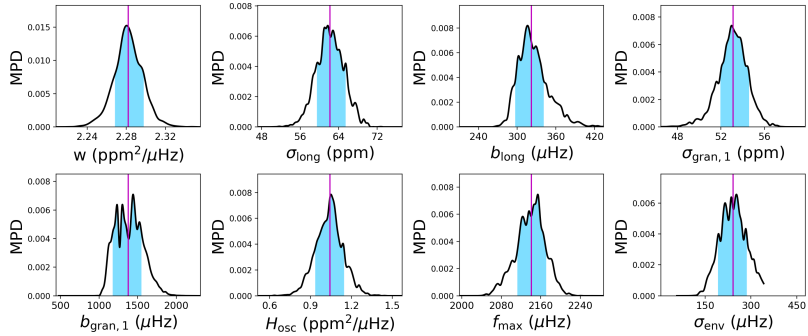

**Figure 10.** Posteriors for the granulation background fit using DIAMONDS (Corsaro & De Ridder 2014). Marginal Probability Distribution (MPD) is plotted against the various fit parameters, with the median value indicated by the vertical line.

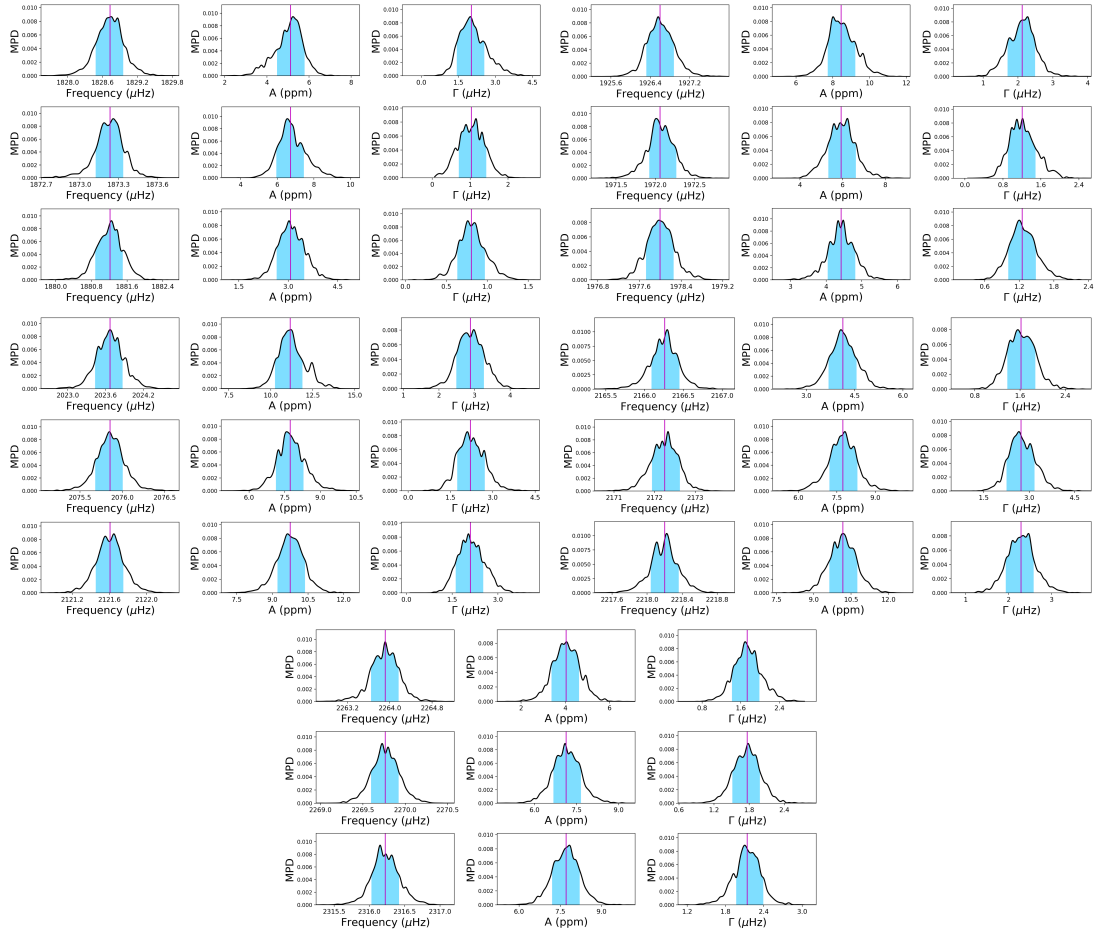

**Figure 11.** Posteriors for the PeakBagging frequency modeling fit using DIAMONDS (Corsaro & De Ridder 2014). Marginal Probability Distribution (MPD) is plotted against the various fit parameters, with the median value indicated by the vertical line.